\documentclass[aps,pra,redraft,reprint,groupedaddress,showpacs,showkeys]{revtex4-1}
\usepackage{watermark}
 \usepackage{amsmath}
 \usepackage{amsfonts}
 \usepackage{graphicx}
 \usepackage{epsfig}
 \usepackage{epstopdf}
 
 \usepackage{verbatim}

 \usepackage{hyperref}
 \hypersetup{
    bookmarks=true,         
    unicode=false,          
    pdftoolbar=true,        
    pdfmenubar=true,        
    pdffitwindow=false,     
    pdfstartview={FitH},    
    pdftitle={GamSpecComment},   
    pdfauthor={D. G. Green},     
    pdfsubject={GamSpecComment},   
    pdfcreator={D. G. Green},   
    pdfproducer={Producer}, 
    pdfkeywords={keywords}, 
    pdfnewwindow=true,      
    colorlinks=true,       
    linkcolor=blue,          
    citecolor=blue,        
    filecolor=magenta,      
    urlcolor=blue           
}

\begin{document}
\title{Comment on ``Gamma-ray spectra from low-energy positron annihilation processes in molecules" }%
\author{D.~G. Green}%
\email{d.green@qub.ac.uk}%
\author{G.~F. Gribakin}%
\email{g.gribakin@qub.ac.uk}
\affiliation{School of Mathematics and Physics, Queen's University Belfast, BT7\,1NN, Northern Ireland, United Kingdom}%
\date{\today}
\begin{abstract}
In the article by Ma~\emph{et al.}~[Phys.~Rev.~A {\bf 94}, 052709 (2016)], $\gamma$-ray spectra for positron annihilation on molecules were calculated in the independent-particle approximation with the positron wavefunction set to unity. Based on comparisons with experimental data they concluded that inner valence electrons play a dominant role in positron annihilation. These conclusions are incorrect and resulted from fallacious analysis that ignored the known effect of the positron wavefunction on the spectra.
\end{abstract}
\maketitle
In a recent article \cite{Ma:2016}, Ma~\emph{et al.}~reported results of independent-particle-model calculations of $\gamma$-ray spectra for low-energy positron annihilation on molecules assuming a plane-wave positron wavefunction
that they set to unity. They compared the annihilation spectra thus obtained with experimental data and concluded that ``positrons annihilate predominantly with inner valence electrons, especially the lowest occupied valence orbital electrons rather than the outer valence electrons''.

However, it is known \cite{DGG_molgamma,DGG_molgammashort} that the plane-wave approximation adopted by Ma~\emph{et al.}~artificially broadens the $\gamma$-ray spectra. This approximation totally ignores the strong positron repulsion from the atomic nuclei. Consequently, it overestimates the contributions of small distances where electrons move fast, which result in large Doppler shifts of the annihilation $\gamma$-rays. Inclusion of nuclear repulsion in the positron wavefunction is crucial for obtaining accurate spectra for positron annihilation on molecules. Ma~\emph{et al.}~make no reference to \cite{DGG_molgamma,DGG_molgammashort} and ignore the conclusions therein. 
As a result, their analysis of the different electron orbital contributions to the $\gamma$-ray spectra is fallacious and their conclusions are incorrect.

In Sec.~III A of their paper, Ma~\emph{et al.}~also applied their method to calculate the annihilation spectra of noble-gas atoms and claimed that ``the inner valence electrons would dominate the annihilation process''. This is in sharp contradiction with recent high-quality many-body theory calculations \cite{DGreen:2015:core} that fully accounted for the positron interaction with the atom, including the nuclear repulsion and very important electron-positron correlation effects. These calculations (see also \cite{DGG_posnobles}) provided an accurate and essentially complete picture of positron interaction with noble-gas atoms, and showed excellent agreement between the theoretical results and measured spectra for Ar, Kr and Xe \cite{PhysRevLett.79.39}. In particular, the relative contributions of various atomic orbitals to the spectra are now firmly established, leaving no room for speculation.

Finally, the paper by Ma~\emph{et al}.~propagates the notion of ``positrophilic sites'' or ``positrophilic electrons''. This idea of preferential positron annihilation with specific molecular electrons is again based on the fallacious analysis of the spectra obtained using the unit positron wavefunction. It contradicts the accumulated understanding of positron annihilation in atoms and molecules, from both experimental and theoretical studies
(see, e.g., \cite{DGG_molgamma,DGG_molgammashort,DGreen:2015:core,DGG_posnobles,PhysRevA.49.R3147,PhysRevLett.79.39,PhysRevA.55.3586}). To quote some of the earlier papers, measurements of the $\gamma$-ray spectra for fluorocarbons ``suggests that positrons annihilate with equal probability on any valence electron" \cite{PhysRevA.55.3586}, while calculations for hydrocarbons suggest that ``most valence molecular orbitals have comparable annihilation probabilities" \cite{PhysRevA.49.R3147}.  

The calculation of accurate $\gamma$-spectra for positron annihilation on molecules is an important problem that warrants attention. 
Plane-wave approximation calculations do not provide accurate annihilation rates nor spectra. Realistic calculations require proper account of the positron wavefunction. 
Ma \emph{et al.}~acknowledge this in the self-contradictory closing paragraph of their paper.\\[-2ex]

\noindent{\bf  Acknowledgements}.~ DGG is supported by the UK Engineering and Physical Sciences Research Council. 

%

\end{document}